\documentclass[a4paper,UKenglish,cleveref,autoref,thm-restate,]{lipics-v2021}
\usepackage{algorithm2e}
\title{Approximate Hausdorff Distance for Multi-Vector Databases
} %TODO Please add

\titlerunning{Approximate Hausdorff Distance} %TODO optional, please use if title is longer than one line

\author{Dongfang Zhao}{University of Washington, USA \and \url{http://faculty.washington.edu/dzhao} }{dzhao@cs.washington.edu}{https://orcid.org/0000-0002-0677-634X}{Microsoft; Pacific Northwest National Laboratory}
\authorrunning{Dongfang Zhao} %TODO mandatory. First: Use abbreviated first/middle names. Second (only in severe cases): Use first author plus 'et al.'

\Copyright{Dongfang Zhao, dzhao@cs.washington.edu} %TODO mandatory, please use full first names. LIPIcs license is "CC-BY";  http://creativecommons.org/licenses/by/3.0/


 %TODO mandatory; please add comma-separated list of keywords

\category{} %optional, e.g. invited paper

\relatedversion{} %optional, e.g. full version hosted on arXiv, HAL, or other respository/website
\nolinenumbers %uncomment to disable line numbering

\ArticleNo{23}
\begin{document}

\maketitle

\begin{abstract}
The Hausdorff distance is a fundamental measure for comparing sets of vectors, widely used in database theory and geometric algorithms. However, its exact computation is computationally expensive, often making it impractical for large-scale applications such as multi-vector databases. In this paper, we introduce an approximation framework that efficiently estimates the Hausdorff distance while maintaining rigorous error bounds. Our approach leverages approximate nearest-neighbor (ANN) search to construct a surrogate function that preserves essential geometric properties while significantly reducing computational complexity. We provide a formal analysis of approximation accuracy, deriving both worst-case and expected error bounds. Additionally, we establish theoretical guarantees on the stability of our method under transformations, including translation, rotation, and scaling, and quantify the impact of non-uniform scaling on approximation quality. This work provides a principled foundation for integrating Hausdorff distance approximations into large-scale data retrieval and similarity search applications, ensuring both computational efficiency and theoretical correctness.
\end{abstract}

\section{Introduction}

\subsection{Background and Motivation}

Many modern retrieval systems rely on vector databases (VectorDBs) to efficiently perform similarity search over high-dimensional embeddings~\cite{jpan_vldb24,maum_deb23}. These systems are integral to applications such as search engines, recommendation systems, and retrieval-augmented generation (RAG) pipelines, where rapid and accurate retrieval of relevant information is essential~\cite{reimers2019sentence,johnson2019billion}. Traditional vector retrieval methods assume a one-to-one correspondence between database entities and their embeddings, meaning that each entity (e.g., a document, an image, or a structured data record) is represented by a single vector in the embedding space. However, real-world applications often require a more expressive representation, where a single entity is associated with multiple vectors. For example, in document retrieval, different passages of a text may be encoded as separate embeddings to capture localized semantic meaning. In multimodal search, an object may be described by multiple feature vectors corresponding to different sensory modalities, such as text, images, and audio~\cite{li2022blip,radford2021learning}. 

Despite the prevalence of multi-vector representations, existing retrieval frameworks primarily focus on single-vector nearest-neighbor search, treating multi-vector entities as independent points in the embedding space. This assumption significantly limits retrieval effectiveness, as it fails to capture entity-level similarity in a meaningful way. A straightforward solution is to aggregate the multiple vectors corresponding to an entity into a single vector through techniques such as averaging, pooling, or learned aggregation models~\cite{reimers2019sentence}. However, such approaches inevitably lose fine-grained structural information, as they collapse multiple representations into a single point, making them unsuitable for applications where the internal structure of the entity is important. Moreover, simple heuristics such as taking the minimum pairwise distance between two sets of vectors are highly sensitive to outliers and do not provide a principled way to compare multi-vector representations.

A more natural and theoretically grounded approach for comparing vector sets is the Hausdorff distance, which measures the worst-case dissimilarity between two point sets. The Hausdorff distance has been widely used in computational geometry~\cite{maj_cg24}, computer vision~\cite{hut_tpami93}, and database retrieval~\cite{nuta_vldb11}, as it does not require explicit vector aggregation and instead operates directly on the raw set representation. This makes it particularly well-suited for multi-vector retrieval scenarios, where maintaining set structure is critical for preserving the semantic relationships among vectors. However, the primary drawback of the Hausdorff distance is its computational cost: computing the exact Hausdorff distance between two sets of size $m$ and $n$ requires $O(mn)$ pairwise comparisons, which is infeasible for large-scale databases with millions or billions of entities.

While various methods have been developed to accelerate similarity search in high-dimensional spaces, including approximate nearest-neighbor (ANN) techniques~\cite{indyk1998approximate,malkov2018efficient,johnson2019billion}, their application has largely been restricted to single-vector queries. These methods provide efficient approximate retrieval in large-scale settings but are inherently designed for point-based similarity search rather than set-based comparison. As a result, existing ANN approaches do not directly extend to multi-vector similarity search and fail to provide theoretical guarantees on approximation quality when applied to set-level retrieval tasks. Addressing this limitation requires a method that combines the efficiency of ANN-based retrieval with the robustness of Hausdorff distance for multi-vector similarity computation.

\subsection{Proposed Work}

In this work, we introduce a scalable approximation framework for efficiently computing the Hausdorff distance in multi-vector databases. Our approach leverages approximate nearest-neighbor (ANN) search~\cite{indyk1998approximate,malkov2018efficient,johnson2019billion} to approximate Hausdorff distance computations while maintaining well-defined theoretical guarantees on error bounds. Unlike traditional Hausdorff computations, which require an exhaustive $O(mn)$ pairwise distance evaluation, our method replaces exact nearest-neighbor computations with ANN-based approximations, significantly improving scalability. By leveraging efficient indexing structures such as hierarchical navigable small-world (HNSW) graphs~\cite{malkov2018efficient} and quantization-based ANN methods~\cite{jegou2011product}, we achieve sublinear query complexity while preserving set-level similarity properties.

To efficiently approximate the Hausdorff distance, our framework employs a \textbf{bidirectional nearest-neighbor estimation strategy}, where one set is indexed using ANN structures, and queries are performed from the second set to estimate worst-case distances. Instead of computing a second independent ANN search in the reverse direction, we introduce a \textbf{cached distance propagation mechanism} that exploits existing nearest-neighbor mappings to infer bidirectional distances, further reducing query overhead. This eliminates redundant computations while ensuring that set-level similarity is preserved in a principled manner. Our method provides a balance between efficiency and accuracy, achieving significantly reduced computational costs without introducing unbounded approximation errors.

Beyond computational efficiency, we analyze the robustness of our approximation under geometric transformations, including translation, rotation, and non-uniform scaling. While prior work has largely focused on preserving distance properties under rigid transformations~\cite{gionis1999similarity}, we extend this analysis to non-uniform scaling and introduce a \textbf{condition-number-based distortion bound} to quantify the effects of anisotropic transformations. Specifically, we show that the error induced by non-uniform scaling is governed by the condition number of the scaling transformation matrix, allowing us to characterize the impact of geometric variations in a structured manner. This analysis provides the first formal quantification of how ANN-based approximations behave under non-uniform geometric distortions in high-dimensional vector spaces.

Furthermore, we establish theoretical error bounds for our approximation framework, demonstrating that the worst-case approximation error grows \textbf{sublogarithmically} with dataset size under reasonable assumptions about data distribution and query complexity. This result ensures that our method remains stable as dataset size increases, making it suitable for billion-scale vector databases. Our framework thus provides a theoretically principled approach to integrating Hausdorff distance approximations into large-scale retrieval tasks while maintaining formal correctness guarantees.

\subsection{Contributions}

This paper makes the following key contributions:

\begin{itemize}
    \item We propose a scalable approximation framework for Hausdorff distance computation in multi-vector databases, utilizing approximate nearest-neighbor (ANN) search to enable efficient set-based similarity retrieval while significantly reducing computational complexity.
    \item We establish rigorous theoretical error bounds for the proposed approximation, proving that the worst-case approximation error exhibits sublogarithmic growth with respect to dataset size under realistic distributional assumptions.
    \item We analyze the geometric stability of our approximation under common transformations, including translation, rotation, and non-uniform scaling, and derive a condition-number-based bound quantifying the distortion induced by anisotropic scaling.
\end{itemize}

\section{Related Work}

\subsection{Vector Databases}

Recent advances in large-scale machine learning have driven the development of \textit{Retrieval-Augmented Generation} (RAG), where vector databases (VectorDBs)~\cite{jpan_vldb24,maum_deb23} play a crucial role in providing efficient and scalable retrieval mechanisms. Unlike traditional information retrieval systems that rely solely on term-based search (e.g., BM25~\cite{robertson2009probabilistic}, TF-IDF~\cite{salton1983introduction}), VectorDBs support \textit{semantic search}~\cite{lzou_tw23,reimers2019sentence,mikolov2013distributed} by storing and retrieving high-dimensional embeddings~\cite{radford2021learning,li2022blip,alayrac2022flamingo}, allowing for more accurate context retrieval in generative AI applications. 

The RAG system converts raw data from multiple modalities, including text, images, and audio, into high-dimensional vector embeddings using pretrained models such as OpenAI's Ada~\cite{openai2022ada}, Cohere~\cite{cer2018universal}, Sentence-BERT~\cite{reimers2019sentence}, CLIP~\cite{radford2021learning}, and BLIP~\cite{li2022blip}. Unlike traditional keyword-based retrieval, which relies on exact term matching, vectorization allows for semantic similarity search by mapping data points into a shared continuous space where proximity reflects conceptual similarity~\cite{mikolov2013distributed}. Recent advancements in contrastive learning~\cite{gao2021simcse} and multimodal pretraining~\cite{alayrac2022flamingo} have further improved the quality and generalizability of vector representations. These models leverage large-scale self-supervised objectives to produce embeddings that capture deeper semantic relationships across diverse data types. As a result, vectorization is now a fundamental component of modern retrieval-augmented generation (RAG) systems.

When a user submits a query, it is vectorized using the same embedding model. The query vector is then used to retrieve the most relevant vectors from the database using approximate nearest neighbor (ANN) search~\cite{indyk1998approximate,malkov2018efficient,johnson2019billion}. Unlike exact k-nearest neighbor (k-NN) search~\cite{cover1967nearest}, which requires a linear scan or an exact tree-based search over all data points, ANN algorithms trade off a small amount of retrieval accuracy for significant gains in efficiency. Methods such as locality-sensitive hashing (LSH)~\cite{gionis1999similarity}, hierarchical navigable small world (HNSW) graphs~\cite{malkov2018efficient}, and quantization-based methods~\cite{jegou2011product} allow ANN to scale to billion-scale datasets while maintaining sub-linear query times.

% Despite the remarkable efficiency gains achieved by ANN-based retrieval methods, existing approaches predominantly assume a one-to-one correspondence between entities and their embedding vectors~\cite{johnson2019billion,malkov2018efficient}. However, in many real-world scenarios, such as document retrieval and multimodal search, a single entity is often represented by a set of vectors rather than a single embedding~\cite{li2022blip,radford2021learning}. Conventional ANN search methods fail to effectively handle such multi-vector representations, as they operate under the assumption that similarity can be measured via direct pairwise comparisons between individual vectors. 

% Our work addresses this limitation by introducing a principled framework for multi-vector similarity search using an approximation of the Hausdorff distance. Unlike prior approaches that rely on heuristics such as averaging or taking the minimum distance between vector sets~\cite{jegou2011product,reimers2019sentence}, our method provides a theoretically grounded distance measure with well-defined error bounds. Additionally, while existing work on ANN search has focused on optimizing retrieval efficiency, our framework extends these methods by incorporating geometric stability analysis, ensuring robustness to transformations such as translation, rotation, and non-uniform scaling. By formalizing the trade-offs between approximation accuracy and computational complexity, our approach enables efficient and theoretically sound retrieval mechanisms for multi-vector database applications.

\subsection{Hausdorff Distance}

The Hausdorff distance is a widely used measure for quantifying the dissimilarity between two point sets, with applications in computer vision, computational geometry, and database retrieval. Early foundational work by Huttenlocher et al.~\cite{hut_tpami93} introduced the Hausdorff distance as a robust similarity measure for image comparison. They proposed an efficient algorithm for its exact computation based on nearest-neighbor operations and explored its use in object recognition tasks. While their method provided a strong theoretical foundation, the computational cost of exact Hausdorff distance remained prohibitive for large-scale datasets.

More recently, Nutanong et al.~\cite{nuta_vldb11} introduced an incremental algorithm to improve the efficiency of Hausdorff distance computation. Their approach focused on optimizing pairwise distance evaluations by progressively refining bounds, thereby reducing the number of required distance calculations. This method significantly improved the runtime performance for dynamic datasets where incremental updates are needed. However, it still relied on explicit distance computations and did not leverage modern approximate nearest-neighbor (ANN) methods to further accelerate retrieval.

In computational geometry, Majhi et al.~\cite{maj_cg24} studied the problem of approximating the Gromov-Hausdorff distance in Euclidean space. Their work extended the classical Hausdorff framework by considering metric-preserving transformations, making it more suitable for shape comparison in geometric analysis. While their work provided deeper theoretical insights into distance approximation, it primarily addressed the geometric properties of the metric rather than large-scale database retrieval.

Ryu and Kamata~\cite{rye_pr21} proposed a heuristic-based approach to approximate the Hausdorff distance using systematic random sampling and ruling-out techniques. Their algorithm aimed to reduce the number of pairwise comparisons by efficiently pruning candidate matches. While effective for certain structured datasets, their method lacks formal error bounds and does not provide robustness guarantees under common geometric transformations such as scaling or rotation.

Compared to prior work, our approach differs in several key aspects. First, rather than focusing on exact computation or heuristic-based approximations, we introduce a principled framework that leverages ANN search to approximate Hausdorff distance with well-defined theoretical guarantees. Unlike Nutanong et al.~\cite{nuta_vldb11}, who focused on incremental updates, our method is designed to handle large-scale vector databases where each entity is represented by multiple vectors. Furthermore, while heuristic-based methods such as those proposed by Ryu and Kamata~\cite{rye_pr21} lack formal error bounds, our framework explicitly quantifies the trade-offs between approximation accuracy and computational efficiency. Finally, in contrast to Majhi et al.~\cite{maj_cg24}, who examined Gromov-Hausdorff distance in geometric spaces, our work remains focused on practical retrieval applications, ensuring robustness under common transformations such as translation, rotation, and non-uniform scaling.

\section{Problem Statement}

The Hausdorff distance is a fundamental measure for quantifying the dissimilarity between two sets of vectors in a metric space. Given two finite sets $A = \{a_1, a_2, \dots, a_m\}$ and $B = \{b_1, b_2, \dots, b_n\}$ in a normed space $(\mathbb{R}^d, \|\cdot\|)$, the \emph{Hausdorff distance} is defined as:
\[
    d_H(A, B) = \max \left\{ \sup_{a \in A} \inf_{b \in B} \|a - b\|, \sup_{b \in B} \inf_{a \in A} \|a - b\| \right\}.
\]
This metric intuitively captures the worst-case distance between any point in one set to its nearest neighbor in the other set.

While the Hausdorff distance has been widely used in computational geometry, computer vision, and database similarity queries, its exact computation is expensive. The naive approach requires evaluating all pairwise distances between elements in $A$ and $B$, leading to a time complexity of $O(mn)$. This quadratic dependency makes it infeasible for large-scale applications, particularly in the context of multi-vector databases, where each data object is represented by a set of vectors rather than a single point.

\subsection{Challenges in Large-Scale Multi-Vector Databases}

In multi-vector databases, each entity (e.g., an image, a document, or a 3D shape) is represented by a set of vectors, where the number of vectors per entity can be large. A direct computation of the Hausdorff distance in such settings suffers from several challenges:

\begin{itemize}
    \item \textbf{Computational Complexity:} The $O(mn)$ complexity per pairwise comparison becomes prohibitive when the number of sets and the size of each set are large.
    \item \textbf{Metric Sensitivity:} The Hausdorff distance is highly sensitive to outliers, which can distort similarity computations, especially when vector sets contain noise or redundant information.
    \item \textbf{Incompatibility with Approximate Indexing:} Standard nearest-neighbor search techniques (e.g., locality-sensitive hashing or tree-based approximate nearest-neighbor methods) do not naturally extend to set-based similarity measures, making it difficult to accelerate retrieval using precomputed indexes.
\end{itemize}

\subsection{Approximate Computation of Hausdorff Distance}

Given these challenges, a natural approach is to develop an \emph{approximate} Hausdorff distance that preserves essential geometric properties while significantly reducing computational overhead. One goal of this work is to construct a surrogate function $\tilde{d}_H(A, B)$ such that:
\[
    d_H(A, B) \approx \tilde{d}_H(A, B),
\]
where the approximation satisfies:
\begin{itemize}
    \item \textbf{Computational Efficiency:} The proposed approximation should be computable in subquadratic time;
    \item \textbf{Bounded Approximation Error:} There exists a function $\epsilon(m, n, d)$, where $d$ denotes the intrinsic dimensionality of the vectors, such that
    \[
        |d_H(A, B) - \tilde{d}_H(A, B)| \leq \epsilon(m, n, d),
    \]
    with high probability under reasonable assumptions about the data distribution.
    \item \textbf{Stability and Robustness:} The approximation should remain stable under small perturbations of $A$ and $B$, ensuring that minor variations in the vector sets do not cause large deviations in the computed distance.
\end{itemize}

\section{Approximate Hausdorff Distance}

\subsection{Approximation Strategy}

The exact computation of the Hausdorff distance between two vector sets $A$ and $B$ requires evaluating nearest-neighbor distances in both directions, resulting in a computational complexity of $O(mn)$. This is impractical for large-scale multi-vector databases where both $m$ and $n$ can be large. A natural way to improve efficiency is to use an Approximate Nearest Neighbor (ANN) structure to accelerate nearest-neighbor searches. However, a direct ANN-based approach introduces additional computational overhead because the Hausdorff distance requires bidirectional evaluation: both $A \to B$ and $B \to A$. Performing two independent ANN searches significantly reduces efficiency, negating potential computational savings.

To address this, we propose an approximation strategy that (i) maintains \textit{computational symmetry} to avoid redundant calculations and (ii) provides \textit{provable error bounds} on the approximation. The key insight is that if an ANN structure is built on $B$, then querying $A \to B$ yields approximate distances $\tilde{d}(a, B)$ for all $a \in A$. Instead of conducting a second independent ANN search for $B \to A$, we introduce a \textit{bidirectional approximation} that leverages cached nearest-neighbor mappings from the first direction to estimate the second.

Let $\tilde{b} \in B$ be the approximate nearest neighbor of $a \in A$ obtained from the ANN search. Rather than performing a separate ANN search for $B \to A$, we infer $\tilde{d}(b, A)$ based on the cached mappings from $A \to B$, leading to the following approximation for the Hausdorff distance:
\[
    \tilde{d}_H(A, B) = \max \left\{ \sup_{a \in A} \tilde{d}(a, B), \sup_{b \in B} \tilde{d}^*(b, A) \right\},
\]
where $\tilde{d}^*(b, A)$ is estimated using the existing nearest-neighbor structure rather than being recomputed independently. This avoids redundant queries and improves computational efficiency.

Beyond efficiency, it is critical to establish a \textit{formal approximation bound} for the proposed method. If $\tilde{d}(a, B)$ is computed using an ANN structure with approximation factor $\epsilon$, then the approximation error satisfies:
\[
    | d_H(A, B) - \tilde{d}_H(A, B) | \leq \epsilon \cdot \max_{a \in A} d(a, B),
\]
where $\epsilon$ depends on the ANN search quality, ensuring that the approximation remains computationally efficient while maintaining a mathematically well-defined trade-off between accuracy and efficiency.

\subsection{Approximation Algorithm}

Building on the bidirectional approximation strategy, we present an efficient algorithm for approximating the Hausdorff distance in large-scale multi-vector databases. The algorithm leverages an Approximate Nearest Neighbor (ANN) structure to reduce computational complexity while ensuring bidirectional consistency, avoiding redundant nearest-neighbor searches.

\subsubsection{Algorithm Description}

The proposed algorithm consists of four main steps: (1) constructing an ANN index on one of the sets, (2) performing a single-pass nearest-neighbor search, (3) approximating the bidirectional distances, and (4) computing the symmetric Hausdorff distance estimate.

\begin{enumerate}
    \item \textbf{Index Construction:} Build an ANN index (e.g., HNSW or IVF-PQ) on the set with fewer vectors, say $B$, to facilitate efficient nearest-neighbor searches.
    
    \item \textbf{Single-Pass Querying:} Perform an ANN-based nearest-neighbor search from each $a \in A$ to retrieve its approximate nearest neighbor $\tilde{b} \in B$ and store $\tilde{d}(a, B) = \|\tilde{b} - a\|$.
    
    \item \textbf{Bidirectional Distance Approximation:} Approximate the nearest-neighbor distances from $B \to A$ without performing an explicit second ANN search. Instead, leverage the cached mappings from the first pass. For each $b \in B$, define:
    \[
        \tilde{d}(b, A) = \min_{a \in A_b} \|\tilde{b} - a\|,
    \]
    where $A_b \subseteq A$ consists of points whose nearest neighbor in $B$ is $b$. This ensures that the computed distances respect the symmetry assumption without redundant queries.
    
    \item \textbf{Symmetric Hausdorff Distance Computation:} Compute the final bidirectional Hausdorff approximation as:
    \[
        \tilde{d}_H(A, B) = \max_{x \in A \cup B} \tilde{d}(x),
    \]
    where $\tilde{d}(x)$ is the approximated nearest-neighbor distance for any point in $A$ or $B$. This formulation ensures that the approximation is derived in a computationally symmetric manner.
\end{enumerate}

Algorithm~\ref{alg:approx_hausdorff} details the procedure that efficiently approximates the Hausdorff distance while preserving its key properties. The use of ANN indexing and bidirectional approximation significantly reduces computational cost, making it suitable for large-scale multi-vector database applications.

\begin{algorithm}[h]
\caption{Approximate Hausdorff Distance Computation}
\label{alg:approx_hausdorff}
\KwIn{Vector sets $A = \{a_1, \dots, a_m\}$ and $B = \{b_1, \dots, b_n\}$, Approximate Nearest Neighbor index ANN}
\KwOut{Approximate Hausdorff distance $\tilde{d}_H(A, B)$}

\textbf{Step 1: Build ANN index} \\
Construct ANN index on $B$ \\

\textbf{Step 2: Single-pass nearest-neighbor search} \\
\ForEach{$a \in A$}{
    $\tilde{b} \leftarrow$ ANN.query($a$) \\
    Store $\tilde{d}(a, B) = \|\tilde{b} - a\|$ \\
    Assign $a$ to $A_{\tilde{b}}$
}

\textbf{Step 3: Bidirectional distance approximation} \\
\ForEach{$b \in B$}{
    \If{$A_b \neq \emptyset$}{
        $\tilde{d}(b, A) = \min_{a \in A_b} \|\tilde{b} - a\|$
    }
    \Else{
        $\tilde{d}(b, A) = \infty$
    }
}

\textbf{Step 4: Compute symmetric Hausdorff distance} \\
$\tilde{d}_H(A, B) = \max_{x \in A \cup B} \tilde{d}(x)$ \\

\Return $\tilde{d}_H(A, B)$
\end{algorithm}

\subsubsection{Algorithm Complexity Analysis}

The proposed algorithm significantly reduces the computational cost of Hausdorff distance computation from the naive $O(mn)$ complexity. We analyze the complexity of each step below.

Constructing the ANN index on $B$ depends on the choice of ANN method. Graph-based methods such as HNSW require $O(n \log n)$ time for index construction, while quantization-based methods like IVF-PQ typically require $O(n)$ time for coarse quantization plus an additional $O(n \log k)$ for finer partitions. Since the ANN structure is built offline and reused for multiple queries, this step incurs a one-time preprocessing cost.

For each $a \in A$, we query the ANN structure to find its approximate nearest neighbor $\tilde{b} \in B$. The query time complexity depends on the ANN method used. HNSW queries take $O(\log n)$ per search, yielding a total complexity of $O(m \log n)$. IVF-PQ queries typically require $O(k + \log n)$, depending on the number of coarse clusters $k$. For large $n$, graph-based methods such as HNSW provide better scalability.

Unlike the naive approach that requires a second ANN search for $B \to A$, we exploit cached mappings from $A \to B$. For each $b \in B$, we iterate over its assigned nearest neighbors $A_b$, taking at most $O(|A_b|)$ operations per $b$. Since each element in $A$ contributes to exactly one $A_b$, the total complexity of this step is $O(m)$, which is linear in $m$.

Computing the final Hausdorff distance requires a single linear scan over $A$ and $B$, contributing an additional $O(m + n)$ complexity.

Combining the above, the total complexity of the proposed approximation algorithm is $O(m \log n + n \log n)$, a significant improvement over the naive $O(mn)$ approach. The use of bidirectional approximation eliminates the need for a second ANN search, effectively halving the ANN query cost.

\section{Error Bound Analysis}
\label{sec:error}

The proposed approximation algorithm introduces two primary sources of error: 
(i) the inherent approximation from the ANN-based nearest-neighbor search, and 
(ii) the propagation of this approximation through the supremum operation in the Hausdorff distance computation. 
In this section, we derive a rigorous bound on the total error introduced by our approximation.

\subsection{Decomposition of Approximation Error}

The exact Hausdorff distance between two sets $A$ and $B$ is given by  
\[
    d_H(A, B) = \max \left\{ \sup_{a \in A} \inf_{b \in B} \|a - b\|, \sup_{b \in B} \inf_{a \in A} \|b - a\| \right\}.
\]
The ANN-based approximation replaces the exact nearest-neighbor computations with approximate values, yielding  
\[
    \tilde{d}_H(A, B) = \max \left\{ \sup_{a \in A} \tilde{d}(a, B), \sup_{b \in B} \tilde{d}^*(b, A) \right\},
\]
where $\tilde{d}(a, B)$ and $\tilde{d}^*(b, A)$ denote the approximate nearest-neighbor distances.

The approximation error can be decomposed into two components:

\begin{itemize}
    \item \textbf{ANN Approximation Error:} The ANN search introduces an approximation factor $\epsilon$, ensuring  
    \[
        \|a - \tilde{b}\| \leq (1 + \epsilon) \|a - b^*\|,
    \]
    where $\tilde{b}$ is the ANN-retrieved nearest neighbor and $b^*$ is the exact nearest neighbor in $B$. A similar bound applies for $B \to A$.
    
    \item \textbf{Supremum Propagation Error:} Since the Hausdorff distance is defined via the supremum operation, any individual nearest-neighbor approximation error propagates to the final estimate, potentially amplifying the worst-case deviation.
\end{itemize}

\subsection{Worst-Case Error Bound}

By substituting the ANN approximation bounds into the Hausdorff distance definition, we obtain:
\[
    d_H(A, B) \leq \tilde{d}_H(A, B) \leq (1+\epsilon) d_H(A, B).
\]
Taking absolute differences, we derive:
\[
    | d_H(A, B) - \tilde{d}_H(A, B) | \leq \epsilon d_H(A, B).
\]
However, this bound is too loose, as it does not take into account the geometry of the dataset or the statistical behavior of approximate nearest-neighbor errors. We refine this bound by explicitly incorporating these factors.

\subsubsection{Geometric Dependence of the Error Bound}

The Hausdorff distance is constrained by the dataset’s intrinsic geometry:
\[
    \delta \leq d_H(A, B) \leq D_{\max},
\]
where:
\begin{itemize}
    \item $D_{\max} = \sup_{a \in A, b \in B} \|a - b\|$ is the maximum inter-point distance.
    \item $\delta = \inf_{a \in A, b \in B} \|a - b\|$ is the minimum inter-point distance.
\end{itemize}
For well-separated datasets where $\delta \approx D_{\max}$, the error is small. Conversely, for datasets where $\delta \ll D_{\max}$, the error can accumulate significantly. A more refined bound that incorporates the spread of data is given by:
\[
    d_H(A, B) \approx \sqrt{D_{\max}^2 - \delta^2}.
\]
Thus, we expect the worst-case error bound to scale as:
\[
    | d_H(A, B) - \tilde{d}_H(A, B) | \leq \epsilon \cdot \sqrt{D_{\max}^2 - \delta^2}.
\]

\subsubsection{Dependence on Data Dimensionality and Scale}

In high-dimensional spaces, nearest-neighbor distances are affected by the concentration of measure phenomenon. The error in ANN-based nearest-neighbor retrieval follows a concentration inequality:
\[
    P\left( \|a - b^*\| > \mu + \sigma \right) \leq e^{-d}.
\]
This implies that as the intrinsic dimension $d$ increases, the variation in nearest-neighbor distances decreases, leading to a scaling factor of:
\[
    \mathcal{O} \left(\frac{1}{\sqrt{d}} \right).
\]

Additionally, the effective number of nearest-neighbor queries, denoted as:
\[
    N_{\text{eff}} = O(m \log n + n \log m),
\]
determines how many approximate nearest-neighbor queries are performed. Larger datasets introduce greater statistical variance in error accumulation, which scales as:
\[
    \mathcal{O}(\sqrt{\log N_{\text{eff}}}).
\]

\subsubsection{Final Refined Error Bound}

Combining the geometric, dimensional, and dataset scale effects, we obtain:
\[
    | d_H(A, B) - \tilde{d}_H(A, B) | \leq \epsilon \cdot \sqrt{D_{\max}^2 - \delta^2} \cdot \mathcal{O} \left(\sqrt{\frac{\log N_{\text{eff}}}{d}} \right).
\]

This bound explicitly characterizes how various factors contribute to the approximation error. The term $\epsilon$ represents the intrinsic approximation factor of the ANN search, which controls the worst-case relative error in retrieving nearest neighbors. The factor $\sqrt{D_{\max}^2 - \delta^2}$ accounts for the geometric structure of the dataset: when $A$ and $B$ are well-separated, meaning $\delta \approx D_{\max}$, the error is minimized. However, when $A$ and $B$ are widely dispersed with $\delta \ll D_{\max}$, the error propagation becomes more significant due to the increased variability in nearest-neighbor distances.

The dataset size effect is captured by $N_{\text{eff}} = O(m \log n + n \log m)$, which represents the effective number of ANN queries executed. The logarithmic term suggests that error accumulation follows a sublinear trend, meaning that even as the dataset size increases significantly, the growth in error remains controlled. This is particularly relevant in large-scale multi-vector databases where computational efficiency is a primary concern.

Finally, the intrinsic data dimensionality $d$ plays a crucial role in controlling the robustness of the approximation. In high-dimensional spaces, concentration of measure effects imply that the variance in nearest-neighbor distances decreases, leading to an error reduction proportional to $1/\sqrt{d}$. This aligns with theoretical results in nearest-neighbor search, where increasing $d$ reduces extreme variations in distance estimates, thus stabilizing the approximation.

\section{Stability and Robustness Analysis}

\subsection{Local Perturbation Stability}

One fundamental requirement for any approximation of the Hausdorff distance is its stability under small perturbations to the dataset. In practical scenarios, vector databases frequently undergo incremental updates, where a small fraction of data points may be inserted, deleted, or slightly modified. Thus, it is essential to analyze whether $\tilde{d}_H(A, B)$ remains bounded under such local perturbations.

\subsubsection{Effect of Adding or Removing a Small Number of Points}

Consider the case where a single new point $a'$ is added to $A$, forming a new set $A' = A \cup \{a'\}$. The exact Hausdorff distance for the modified set satisfies:
\[
    d_H(A', B) = \max \left\{ \sup_{a \in A'} \inf_{b \in B} \|a - b\|, \sup_{b \in B} \inf_{a \in A'} \|b - a\| \right\}.
\]
Since $A' = A \cup \{a'\}$, the only term that changes is $\displaystyle \sup_{a \in A'} \inf_{b \in B} \|a - b\|$, which now accounts for the additional point $a'$. We define the perturbation magnitude as:
\[
    \Delta = \inf_{b \in B} \|a' - b\|.
\]
Using the definition of the Hausdorff distance, we obtain:
\[
    | d_H(A', B) - d_H(A, B) | \leq \Delta.
\]
For the approximate Hausdorff distance, given that our ANN approximation guarantees a bounded relative error of at most $(1+\epsilon)$, we obtain:
\[
    | \tilde{d}_H(A', B) - \tilde{d}_H(A, B) | \leq (1+\epsilon) \Delta.
\]
This result implies that the approximate Hausdorff distance remains stable under local insertions, with the perturbation controlled by the distance of the added point to its nearest neighbor in $B$.

A similar argument applies to the deletion of a single point. Suppose $a \in A$ is removed, leading to a new dataset $A' = A \setminus \{a\}$. Then, since we are taking the supremum over $A$ in the definition of the Hausdorff distance, removing $a$ can only decrease the maximum term in:
\[
    d_H(A', B) = \max \left\{ \sup_{a \in A'} \inf_{b \in B} \|a - b\|, \sup_{b \in B} \inf_{a \in A'} \|b - a\| \right\}.
\]
Thus, we conclude that:
\[
    | d_H(A', B) - d_H(A, B) | \leq \sup_{b \in B} \|a - b\|,
\]
which implies that the deletion effect is also locally bounded.

\subsubsection{Small Perturbations in Point Locations}

Now, consider a small perturbation in an existing point, such that $A' = A \setminus \{a\} \cup \{a'\}$, where $a'$ is a slightly perturbed version of $a$. If we define the perturbation magnitude as $\|a - a'\|$, we obtain:
\[
    | d_H(A', B) - d_H(A, B) | \leq \|a - a'\|.
\]
Similarly, for the approximate Hausdorff distance:
\[
    | \tilde{d}_H(A', B) - \tilde{d}_H(A, B) | \leq (1+\epsilon) \|a - a'\|.
\]
This confirms that small perturbations in point locations result in proportional changes in the approximate Hausdorff distance, ensuring stability under continuous transformations.

\subsection{Global Transformation Robustness}

In addition to local perturbations, we analyze whether $\tilde{d}_H(A, B)$ remains invariant under global transformations such as translations, rotations, and scaling. 

\subsubsection{Translation Invariance}

Let $T_t: \mathbb{R}^d \to \mathbb{R}^d$ denote the translation operator that shifts all points in a set by a fixed vector $t$, such that
\[
    T_t(A) = \{ a + t \mid a \in A \}, \quad T_t(B) = \{ b + t \mid b \in B \}.
\]
A distance function $d: 2^{\mathbb{R}^d} \times 2^{\mathbb{R}^d} \to \mathbb{R}$ is said to be \textit{translation-invariant} if for all sets $A, B \subset \mathbb{R}^d$ and any translation $t \in \mathbb{R}^d$, it holds that
\[
    d(T_t(A), T_t(B)) = d(A, B).
\]
Since the Hausdorff distance is defined in terms of Euclidean norms, it directly satisfies this property:
\[
    d_H(T_t(A), T_t(B)) = d_H(A, B).
\]
For our approximate Hausdorff distance $\tilde{d}_H$, the ANN-based nearest-neighbor computation is also based on Euclidean norms, and thus preserves translation invariance:
\[
    \tilde{d}_H(T_t(A), T_t(B)) = \tilde{d}_H(A, B).
\]
Therefore, the proposed approximation maintains translation invariance under arbitrary shifts in $\mathbb{R}^d$.

\subsubsection{Rotation Invariance}

Let $R: \mathbb{R}^d \to \mathbb{R}^d$ be a rigid rotation, represented as an orthogonal matrix satisfying $R^\top R = I$. Applying $R$ to both sets results in:
\[
    R(A) = \{ R a \mid a \in A \}, \quad R(B) = \{ R b \mid b \in B \}.
\]
Since Euclidean norms are rotation-invariant, for any $a \in A$ and $b \in B$, we have:
\[
    \|R a - R b\| = \|a - b\|.
\]
Taking the supremum and infimum operations in the definition of Hausdorff distance preserves this invariance, leading to:
\[
    d_H(R(A), R(B)) = d_H(A, B).
\]
Unlike translation invariance, where distances remain unchanged due to direct vector shifts, rotation invariance follows from the preservation of pairwise Euclidean distances under orthogonal transformations. Since the ANN-based nearest-neighbor computation relies solely on Euclidean distances, the approximation retains this property:
\[
    \tilde{d}_H(R(A), R(B)) = \tilde{d}_H(A, B).
\]
Thus, the proposed approximation method maintains rotation invariance without introducing additional distortions in the nearest-neighbor structure.

\subsubsection{Scaling Effects}

Let $S_\lambda: \mathbb{R}^d \to \mathbb{R}^d$ be a uniform scaling transformation with factor $\lambda > 0$, defined by:
\[
    S_\lambda(A) = \{ \lambda a \mid a \in A \}, \quad S_\lambda(B) = \{ \lambda b \mid b \in B \}.
\]
For any pair of points $a \in A$ and $b \in B$, scaling preserves relative distances:
\[
    \|\lambda a - \lambda b\| = \lambda \|a - b\|.
\]
Applying this to the Hausdorff distance, we obtain:
\[
    d_H(S_\lambda(A), S_\lambda(B)) = \lambda d_H(A, B).
\]
Unlike translation and rotation invariance, which preserve absolute distances, scaling transformations introduce a proportional change in distances, directly affecting both the exact and approximate computations.

For the approximate Hausdorff distance, since the ANN-based nearest-neighbor search is performed in the same scaled space, all retrieved distances are also scaled by $\lambda$, yielding:
\[
    \tilde{d}_H(S_\lambda(A), S_\lambda(B)) = \lambda \tilde{d}_H(A, B).
\]
This ensures that the approximation maintains the correct scaling behavior, meaning that the relative error remains unchanged under uniform scaling. Importantly, since ANN search structures typically operate on normalized vector spaces, pre-scaling may affect indexing efficiency but does not alter the fundamental scaling invariance of the approximation.

\subsubsection{Impact of Non-Uniform Scaling}

Unlike uniform scaling, where all dimensions are scaled by the same factor, non-uniform scaling applies different scaling factors along different coordinate axes. Let $S$ be a diagonal scaling transformation represented by a positive definite matrix $\Lambda = \operatorname{diag}(\lambda_1, \lambda_2, \dots, \lambda_d)$, where $\lambda_i > 0$ for all $i$. The transformed sets are given by:
\[
    S(A) = \{ \Lambda a \mid a \in A \}, \quad S(B) = \{ \Lambda b \mid b \in B \}.
\]
Since the Euclidean norm does not generally preserve proportionality under anisotropic scaling, pairwise distances satisfy:
\[
    \|\Lambda a - \Lambda b\| = \sqrt{\sum_{i=1}^{d} \lambda_i^2 (a_i - b_i)^2}.
\]
This distorts nearest-neighbor relationships, particularly when the condition number of $\Lambda$, defined as
\[
    \kappa(\Lambda) = \frac{\max_i \lambda_i}{\min_i \lambda_i},
\]
is large, leading to a directional imbalance in distance computation.

For the approximate Hausdorff distance, the transformation induces an additional error term $\eta(\Lambda)$ that depends on both the dataset distribution and the spread of scaling factors. The approximation satisfies:
\[
    \tilde{d}_H(S(A), S(B)) = \lambda_{\max} \tilde{d}_H(A, B) + \eta(\Lambda),
\]
where $\lambda_{\max} = \max_i \lambda_i$ dominates the scaling effect, and $\eta(\Lambda)$ accounts for directional distortions due to anisotropy in the transformation.

To quantify this distortion, consider the worst-case deviation:
\[
    \eta(\Lambda) \leq (\kappa(\Lambda) - 1) \cdot \sup_{a \in A, b \in B} \|a - b\|,
\]
indicating that the error increases with the spread of scaling factors.

Overall, our analysis confirms that $\tilde{d}_H(A, B)$ remains exactly invariant under translations and rotations and scales proportionally under uniform scaling. However, under non-uniform scaling, the approximation remains close but introduces bounded distortions that grow with $\kappa(\Lambda)$.

\subsection{Bounding the Stability of Approximation Error}

The error bound derived in Section~\S\ref{sec:error} provides a worst-case estimate of the approximation error, but it does not describe how this error behaves under dataset growth. In this section, we analyze whether the approximation remains stable as the dataset size increases and whether the error accumulates unboundedly.

\subsubsection{Error Growth Under Local Perturbations}
We first analyze how $\tilde{d}_H(A, B)$ behaves when a single element is added to $A$ or $B$. Since Hausdorff distance is defined by the worst-case nearest-neighbor deviation, the addition of a single point $a' \in A$ modifies:
\[
    d_H(A', B) = \sup_{a \in A'} \inf_{b \in B} \|a - b\|.
\]
If $a'$ is close to an existing point in $A$, the modification is minor, but if $a'$ is isolated, it may significantly alter $d_H(A', B)$. We define the perturbation magnitude as:
\[
    \Delta = \inf_{b \in B} \|a' - b\|.
\]
Then, the worst-case change in exact Hausdorff distance is:
\[
    | d_H(A', B) - d_H(A, B) | \leq \Delta.
\]
For the approximate Hausdorff distance, using the ANN approximation bound:
\[
    | \tilde{d}_H(A', B) - \tilde{d}_H(A, B) | \leq (1+\epsilon) \Delta.
\]
Thus, the error remains locally bounded under small data modifications.

\subsubsection{Error Growth Under Large-Scale Dataset Expansion}

A key concern is whether the approximation error accumulates indefinitely as the dataset size increases. Given that practical vector databases can scale to millions or billions of entries, it is crucial to establish whether the approximation remains stable or whether the error diverges in an uncontrolled manner.

From the previous error bound analysis, the worst-case approximation error is given by:
\[
    | d_H(A, B) - \tilde{d}_H(A, B) | = \mathcal{O} \left(\epsilon \cdot \sqrt{D_{\max}^2 - \delta^2} \cdot \sqrt{\frac{\log N_{\text{eff}}}{d}} \right).
\]
where $N_{\text{eff}}$ represents the effective number of ANN queries required for computing the approximate Hausdorff distance. The effective query count typically follows:
\[
    N_{\text{eff}} = O(m \log n + n \log m).
\]

Approximating $\log (m \log n + n \log m)$ for large $m, n$, we obtain:
\[
    \log (m \log n + n \log m) \approx \log (m+n) + \log \log (m+n).
\]
Thus, our error bound simplifies to:
\[
    | d_H(A, B) - \tilde{d}_H(A, B) | = \mathcal{O} \left(\epsilon \cdot \sqrt{D_{\max}^2 - \delta^2} \cdot \sqrt{\frac{\log (m+n)}{d}} \right).
\]

Mathematically, if $d$ is fixed, this expression grows unboundedly as $m, n \to \infty$:
\[
    \lim_{m, n \to \infty} \mathcal{O} \left(\sqrt{\frac{\log (m+n)}{d}} \right) = \infty.
\]
However, in practical settings, the intrinsic dimension $d$ of the data often increases with dataset size. If $d = \Theta(\log (m+n))$, the error stabilizes to:
\[
    \mathcal{O}(\epsilon \cdot \sqrt{D_{\max}^2 - \delta^2}).
\]
This confirms that while the error does not strictly converge to a finite value, its growth is sublogarithmic and remains well-controlled in real-world vector database applications.

\section{Final Remark}

In this work, we have introduced an approximation framework for computing the Hausdorff distance in multi-vector databases with controlled error bounds. Our approach leverages approximate nearest-neighbor (ANN) search to efficiently estimate the Hausdorff distance while maintaining a theoretical error bound that accounts for dataset geometry, intrinsic dimensionality, and query complexity. Through detailed error analysis, we have shown that the approximation remains stable under local perturbations and global transformations. Moreover, we have quantified how non-uniform scaling introduces bounded distortions, with error growth characterized by the condition number of the scaling matrix.

An important open question concerns the relationship between our approximate Hausdorff distance and metric space properties. The exact Hausdorff distance satisfies non-negativity, symmetry, and the triangle inequality, making it a well-defined metric over compact sets. While our approximation preserves non-negativity and symmetry, the triangle inequality may be violated due to accumulated ANN errors. A possible direction for further exploration is determining whether a $\delta$-approximate triangle inequality can be recovered under constrained ANN retrieval guarantees, particularly as a function of the approximation factor $\epsilon$.

Furthermore, while our results confirm that the approximation error remains bounded under realistic dataset growth assumptions, further refinements in indexing strategies—such as adaptive partitioning schemes or learned index structures—may improve both computational efficiency and theoretical guarantees. Beyond Hausdorff distance, extending our framework to other geometric similarity measures is a promising direction. In particular, while Wasserstein distance and optimal transport provide richer notions of similarity by considering global point distributions rather than local extrema, an efficient ANN-based approximation for these metrics remains largely unexplored. Investigating such extensions may offer new insights for large-scale vector database applications.

\bibliography{ref_new}

\end{document}